\documentclass[useAMS,usenatbib]{mn2e}
\usepackage[dvipdfm]{hyperref}
\hypersetup{urlcolor=blue, colorlinks=true,citecolor=blue}

\usepackage{amssymb}
\usepackage{amsmath}
\usepackage{caption}

\usepackage[dvips]{graphicx}
\usepackage{float}
\restylefloat{figure}
\restylefloat{table}

\usepackage[a4paper, total={7in, 9in}]{geometry}

\newfloat{video}{thpb}{aux}
\floatname{video}{Video}

\graphicspath{{figures/}}

\newcommand{\coh}[2]{\mathsf{{#1}}_{{#2}}}

\title[Bayesian Inference for Radio Observations]{Bayesian Inference for Radio Observations}

\author[M. Lochner et al.]{Michelle Lochner$^{1,2,3,4}$, Iniyan Natarajan$^5$, Jonathan T.~L. ~Zwart$^{5,6}$, Oleg Smirnov$^{7,8}$, \newauthor Bruce A. Bassett$^{1,2,3}$, Nadeem Oozeer$^{1,8,9}$ and Martin Kunz$^{1,10}$\\
$^1$African Institute for Mathematical Sciences, 6 Melrose Road, Muizenberg, 7945, South Africa\\
$^2$Department of Mathematics and Applied Mathematics, University of Cape Town, Rondebosch, Cape Town, 7700, South Africa\\  
$^3$South African Astronomical Observatory, Observatory Road, Observatory, Cape Town, 7935, South Africa\\  
$^4$Department of Physics \& Astronomy, University College London, London, WC1E 6BT, U.K.\\
$^5$Astrophysics, Cosmology and Gravity Centre (ACGC), Department of Astronomy, \\\hspace{6pt}University of Cape Town, Private Bag X3, Rondebosch 7701, South Africa\\ 
$^6$Department of Physics \& Astronomy, University of the Western Cape, Private Bag X17, Bellville 7535, South Africa\\
$^7$Department of Physics and Electronics, Rhodes University, PO Box 94, Grahamstown, 6140, South Africa\\ 
$^8$SKA South Africa, 3rd Floor, The Park, Park Road, Pinelands, 7405, South Africa\\ 
$^9$Centre for Space Research, North-West University, Potchefstroom 2520, South Africa\\ 
$^{10}$D\'epartement de Physique Th\'eorique and Center for Astroparticle Physics, Universit\'e de Gen\`eve, \\\hspace{6pt}Quai E.\ Ansermet 24, CH-1211 Gen\`eve 4, Switzerland}

\begin{document}

\date{}

\pagerange{\pageref{firstpage}--\pageref{lastpage}} \pubyear{2015}

\maketitle

\label{firstpage}

\begin{abstract}
New telescopes like the Square Kilometre Array (SKA) will push into a new sensitivity regime and expose systematics, such as direction-dependent effects, that could previously be ignored. Current methods for handling such systematics rely on alternating best estimates of instrumental calibration and models of the underlying sky, which can lead to inadequate uncertainty estimates and biased results because any correlations between parameters are ignored. These deconvolution algorithms produce a single image that is assumed to be a true representation of the sky, when in fact it is just one realisation of an infinite ensemble of images compatible with the noise in the data. In contrast, here we report a Bayesian formalism that simultaneously infers both systematics and science. Our technique, Bayesian Inference for Radio Observations (BIRO), determines all parameters directly from the raw data, bypassing image-making entirely, by sampling from the joint posterior probability distribution. This enables it to derive both correlations and accurate uncertainties, making use of the flexible software MeqTrees to model the sky and telescope simultaneously. We demonstrate BIRO with two simulated sets of Westerbork Synthesis Radio Telescope datasets. In the first, we perform joint estimates of 103 scientific (flux densities of sources) and instrumental (pointing errors, beam width and noise) parameters. In the second example, we perform source separation with BIRO. Using the Bayesian evidence, we can accurately select between a single point source, two point sources and an extended Gaussian source, allowing for `super-resolution' on scales much smaller than the synthesised beam.
\end{abstract}


\section{Introduction}
The high sensitivity of the SKA (up to 50 times more sensitive than current instruments \citep{ska}) combined with a relatively cheap antenna design means a far more careful and detailed treatment of systematics will be required to fully exploit this telescope \citep{noordam2}. The current approach to this calibration problem iteratively applies deconvolution methods such as CLEAN \citep{clean}, alternating with sky and instrumental modelling to determine the best-fitting, calibrated image \citep{selfcal,kamezi1,kamezi2,bhatnagar}. This provides only a point estimate of the model parameters which will in general differ from the true parameters due to random noise \citep{ensslin}. 

A more rigorous approach is to infer the science and instrumental parameters simultaneously, deriving accurate uncertainties and correlations between them. Work in this direction includes improvements on the self-calibration algorithm \citep{selfcal,ensslin,ensslin2,dorn} and some extensions to the RESOLVE algorithm \citep{junklewitz2,junklewitz}. There has also been considerable effort in this direction in producing a maximum posterior image for the data and dealing with certain calibration parameters \citep{sutton,sutter,sutter2}. These works each solve specific aspects of the calibration and deconvolution problem, but so far do not explore the full posterior distribution, giving an inaccurate estimation of the uncertainties and correlations, and still rely on producing a single image (i.e. a point estimate). 

We propose instead a new technique, called Bayesian Inference for Radio Observations (BIRO), which is able to: include any source of instrumental uncertainty, such as ionospheric effects, pointing errors and primary beam uncertainties, jointly determine the science and instrumental parameters and provide reliable estimates of the uncertainties and correlations on these parameters, in a holistic and mathematically rigorous manner. 

A simultaneous analysis requires the full posterior probability distribution of the parameters, which can naturally be sampled in the Bayesian formalism by using (for example) MCMC \citep{metropolis,hastings} or nested sampling \citep{skilling}. Our new technique, BIRO, fits models including both instrumental and science parameters directly to the raw visibility data. We use the MeqTrees \citep{noordam} software, which implements the Radio Interferometry Measurement Equation (RIME) \citep{hamaker}, for the modelling of the sky and instrumental effects. This technique thus obviates the need for intermediate imaging and map-making. The rigorous statistical use of all available information allows this technique to open new discovery windows, solving previously intractable problems, and is applicable to all interferometers and problems in radio interferometry. 

This paper is arranged as follows: in section \ref{sec:bayes} we provide an introduction to Bayesian statistics and illustrate the use of the RIME for modelling in the BIRO algorithm in section \ref{sec:rime}. We then apply BIRO to two key simulated datasets to demonstrate its power: In section \ref{sec:joint}, we jointly fit all scientific (source flux densities)  and instrumental parameters (pointing errors, primary beam parameters and receiver noise) to a dataset suffering from direction-dependent instrumental effects. In section \ref{sec:source_sep}, we focus on the problem of reliably distinguishing between an extended source, point source and a pair of close point sources, for sources on sub-synthesised beam scales. We conclude in section \ref{sec:conclusions}.

\section{Bayesian statistics}
\label{sec:bayes}
The problem of obtaining the most information possible from an incomplete dataset, such as obtained by an interferometer, is perfectly suited to the application of Bayesian statistics. These allow the fitting of arbitrarily complex models to data, providing reliable uncertainty estimates for the parameters. Bayes' theorem allows the use of a familiar quantity, the likelihood, to answer the question one is really interested in: what is the probability of an hypothesis, given the data in hand? This probability is known as the posterior and indicates by how much our degree of belief in the hypothesis has been updated by the new data. Simple application of Bayes' theorem also allows a robust and intuitive way to compare models, which we will require for the second example problem in this paper. What follows here is a brief overview of Bayesian theory, see \citet{trotta} for a more in-depth review.

From Bayes' theorem, the probability distribution, $\mathcal{P}\left(\mathbf{\Theta}|\mathrm{\mathbf{D}},H\right)$, of the values of parameters
$\mathbf{\Theta}$, the quantity that is actually sought, given the
data $\mathrm{\mathbf{D}}$ that are in-hand and a model $H$
(hypothesis plus any assumptions), is:
\begin{equation} 
\mathcal{P}\left(\mathbf{\Theta}|\mathrm{\mathbf{D}},H\right)
= \frac{
\mathcal{L}\left(\mathrm{\mathbf{D}}|\mathbf{\Theta},H\right)
\mathit{\Pi}\left(\mathbf{\Theta}| H\right)}
{\mathcal{Z}\left(\mathrm{\mathbf{D}}| H\right)}.
\end{equation} 
This is known as the posterior probability distribution. The likelihood
$\mathcal{L}\left(\mathrm{\mathbf{D}}|\mathbf{\Theta},H\right)$, which
encodes any constraints imposed by observations, is the probability
distribution of the data given parameter values and a model. 

The prior
$\mathit{\Pi}\left(\mathbf{\Theta}| H\right)$ includes any prior
knowledge of or prejudices about the parameter values.
$\mathcal{Z}\left(\mathrm{\mathbf{D}}| H\right)$ is the integral of
$\mathcal{L}\left(\mathrm{\mathbf{D}}|\mathbf{\Theta},H\right)
\mathit{\Pi}\left(\mathbf{\Theta}| H\right)$ over all
$\mathbf{\Theta}$, not simply normalizing the posterior
$\mathcal{P}\left(\mathbf{\Theta}|\mathrm{\mathbf{D}},H\right)$, but
also allowing selection of different models by comparing their
values quantitatively. This so-called evidence, $\mathcal{Z}\left(\mathrm{\mathbf{D}}| H\right)$, automatically includes an Occam's razor effect, penalising models with a large number of parameters that are not preferred by the data. By computing the evidence for a range of models we can select the best model by maximising the evidence.

For this work, the likelihood function is
\begin{equation}
\label{eq:likelihood}
 \mathcal{L}\left(\mathrm{\mathbf{D}}|\mathbf{\Theta},H\right)=\frac{1}{(2\pi\sigma^2)^{N/2}} \, \text{exp}\bigg[ -\bigg(\displaystyle \sum_{i=1}^N \Big(\coh{V_i}{}(\mathbf{\Theta}) - \coh{\widetilde{V}}{i} \Big)^2\bigg)/2\sigma^2 \bigg],
\end{equation}
where $\coh{V_i}{}(\mathbf{\Theta})$ are the model visibilities produced by MeqTrees (see section \ref{sec:rime}), with the parameters $\mathbf{\Theta}$ as input, $\coh{\widetilde{V}}{i}$ are the data visibilities, $N$ is the number of data points. Here we assume the uncertainties on the visibilities are Gaussian and have the same value, $\sigma$, for all datapoints. The best-fitting model corresponds to maximum posterior.

The inferred posterior distributions are full probability distributions rather
than a summary mean/median value and a (perhaps covariant)
uncertainty, since this represents the total inference about the
problem at hand. These distributions may be highly non-Gaussian, making such summary parameters inaccurate.

The application of Bayesian statistics allows one to marginalise out the effects of nuisance parameters, which are parameters such as the beam shape and pointing errors that are not of primary interest, but are unknown and can affect the estimates of the parameters of interest (i.e. science parameters) because of correlations and degeneracies. The marginalised posterior can be written as a function of the parameters of interest, $\mathbf{\Phi}$, the nuisance parameters, $\mathbf{\Psi}$, and the data, $\mathbf{D}$:
\begin{equation}
 P(\mathbf{\Phi}|\mathbf{D},H) = \int P(\mathbf{\Phi},\mathbf{\Psi}|\mathbf{D},H) \text{d}\mathbf{\Psi},
\end{equation}
where the integral is performed over the parameter space of $\mathbf{\Psi}$.

The posterior is, thanks to advances in modern computing, fairly easily determined using numerical techniques. In this paper, we use the Markov Chain Monte Carlo (MCMC) Metropolis-Hastings \citep{metropolis,hastings} algorithm for the joint scientific and instrumental parameter inference example. We chose MCMC due to its simplicity and the ease with which it handles large numbers of parameters (we have 103 parameters for the first example problem). For our second example, that of model selection related to source separation and extended structure, we require efficient calculation of the Bayesian evidence, something provided naturally by the nested sampling algorithm. We utilise the public code MultiNest \citep{multinest1,multinest2} to determine both the parameters and the evidence for model comparison \citep{jeffreys,trotta}, but for a smaller set of parameters as nested sampling grows rapidly in complexity with increasing number of parameters.

\section{Using the RIME for modelling}
\label{sec:rime}
Previous Bayesian visibility analyses \citep{lancaster,feroz,zwart,ami,sutter} focused on the sky model and were not generalised to include arbitrary instrumental effects (or were attempting to solve for a much more general sky model resulting in many more parameters, thus needing to fix instrumental parameters). 
The Radio Interferometry Measurement Equation (RIME) \citep{hamaker,smirnov,smirnov2} provides a powerful framework to easily describe exactly what happens to a signal as it travels from source to telescope, where it is converted into voltages. The RIME is a natural way to model the instrumental and scientific effects that we are inferring through our Bayesian technique. For example, the RIME for a single point source is given by
\begin{equation}
\label{eq:rime}
 \coh{V}{pq} = \boldsymbol{J}_p \coh{B}{} \boldsymbol{J}_q^H.
\end{equation}
where $\coh{B}{}$ is the brightness matrix, which describes the sky flux distribution, $\boldsymbol{J}_p$ is the Jones matrix \citep{jones} for antenna $p$, containing all instrumental and atmospheric effects that interfere with the signal, $\boldsymbol{J}_q$ is the Jones matrix for antenna $q$, $H$ indicates the Hermitian of a matrix and $\coh{V}{pq}$ are the visibilities, the outputs of the telescope correlator for baseline $pq$.

The effects that interfere with the signal on its route to the output of the telescope can each be described by a Jones matrix, with each effect adding a pair of Jones matrices in the `onion' form of the RIME:
\begin{equation}
 \coh{V}{pq} = \boldsymbol{J}_{pn} (\ldots(\boldsymbol{J}_{p2}(\boldsymbol{J}_{p1} \coh{B}{} \boldsymbol{J}_{q1}^H)\boldsymbol{J}_{q2}^H)\ldots)\boldsymbol{J}_{qm}^H.
\end{equation}

We can go a few steps further and consider the full-sky RIME by integrating over the direction cosines, $l$ and $m$:
\begin{equation}
\label{eq:rime_full}
 \coh{V}{pq} = \boldsymbol{G}_p \left( \int \int_{lm} \boldsymbol{E}_p \boldsymbol{K}_p \coh{B}{} \boldsymbol{K}_q^H \boldsymbol{E}_q^H \text{d}l \text{d}m \right)\boldsymbol{G}_q^H.
\end{equation}
Here, $\boldsymbol{K}_p$ and $\boldsymbol{K}_q$ are the Jones matrices describing the geometric delay between antennas $p$ and $q$, $\boldsymbol{G}_p$ represents the \emph{direction-independent} gains for antenna $p$, which we set to unity for all antennas, and $\boldsymbol{E}_p$ is the Jones matrix containing all the \emph{direction-dependent} effects for antenna $p$. We focus in this paper on the more difficult to handle direction-dependent effects, but direction-independent can also be handled with our technique. As with all other Jones matrices, $\boldsymbol{E}_p$ can be written as a product of Jones matrices, each describing a different effect. In section \ref{sec:joint}, we consider both primary beam effects and pointing errors as examples of direction-dependent effects each with their own Jones matrix.

The RIME is implemented in the general, flexible software MeqTrees \citep{noordam,smirnov3,smirnov4}, which allows us to apply it to any sky model and for any telescope. MeqTrees has been useful for predicting the capabilities of future experiments and for understanding the intricacies of current telescopes. Here, we go a step further and use MeqTrees as the modelling step in our Bayesian analysis. In order to test BIRO and compare it with the standard deconvolution approach, we use datasets simulated with MeqTrees over which we have complete control and thus would know if we were correctly recovering the true input parameters.

MeqTrees takes from the user a sky model (such as the number and distribution of sources, their fluxes, shapes etc.) as well as instrumental details (such as the telescope configuration, primary beam pattern, pointing errors, noise, atmospheric effects, ionospheric effects etc.) and uses the measurement equation to produce realistic simulated visibilities that such a telescope would observe. 

In order to test the validity of our technique, we only work with simulations in this paper. We use MeqTrees to simulate the data and also to model the sky, to test if we recover the input parameters. MeqTrees can be used to model any telescope configuration and any sky and instrumental effects that can be described with the RIME. While we only concentrate on primary beam and pointing error effects in section \ref{sec:joint}, in principle, a wide variety of source types and instrumental corruptions can be added in MeqTrees. 

\begin{figure}
  \centering
   \includegraphics[width=1\linewidth]{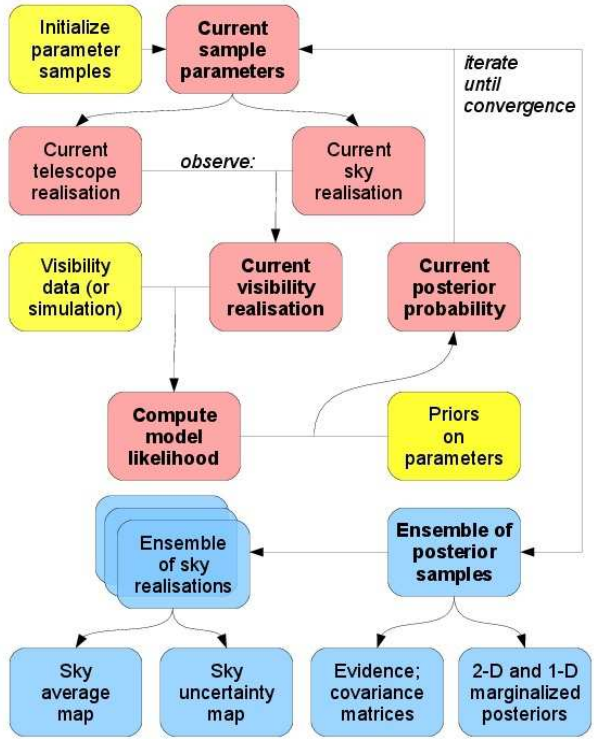}
  \caption{The BIRO algorithm. Fixed or initialized inputs are shown in yellow, while the sampling loop is represented by the pink boxes. Data products are in blue. The main iteration loop occurs within either the MCMC or MultiNest algorithm (depending on the problem), where new parameters are used in each iteration to compute the likelihood. The initial parameters are drawn from the prior, which generally restricts the parameter ranges. In the final step, the ensemble of sky realisations can be generated with MeqTrees using the parameter samples in the posterior, if required.}
  \label{fig:flowchart}
 \end{figure}

Fig. \ref{fig:flowchart} shows a schematic overview of the BIRO approach. At each step in the chain of MCMC or MultiNest, MeqTrees is called with new values for the parameters. MeqTrees then returns a visibility set that can be compared directly with the simulated data, to determine how well the parameters fit. This iterative process allows the determination of the full posterior for the parameters. We do not as yet have a public release of the BIRO code, but plan to in the future where we will integrate MontBlanc, a GPU implementation of the RIME \citep{perkins} with BIRO. MontBlanc is already publicly available meaning it can be combined with any sampler to allow the user to implement BIRO for themselves.

\section{Example 1: Joint inference of scientific and instrumental parameters}
\label{sec:joint}
In this example, we use BIRO to jointly estimate the scientific parameters and nuisance instrumental parameters. Below we describe the model and simulated dataset used, and details of the MCMC analysis, and show that the instrumental parameters studied are tightly correlated with the scientific parameters, a fact that cannot be ignored when determining these parameters.

\subsection{Simulated data and parameters of the model}
\subsubsection{Telescope configuration}
We use MeqTrees to simulate observations with the Westerbork Synthesis Radio Telescope (WSRT) \citep{wsrt}, a 14-element East-West array with 25m diameter dishes. All our WSRT simulations use an integration time of 30 seconds and a total observation time of 12 hours at a frequency of 1.4 GHz. We use a narrow bandwidth of 125kHz, a single channel (for simplicity) and include noise with a standard deviation of 0.1 Jy/visibility. At this frequency, WSRT has a field of view of 0.5-0.6 degrees and a synthesised beam width of around 13 arcsec FWHM (full width at half maximum)\footnote{WSRT Guide to Observations, \url{www.astron.nl/radio-observatory/astronomers/wsrt-guide-observations/5-technical-information/5-technical-informatio}}.

\subsubsection{Scientific parameters}
\begin{figure}
\centering
\includegraphics[width=1\linewidth,trim=0cm 0cm 0cm 0cm, clip=true]{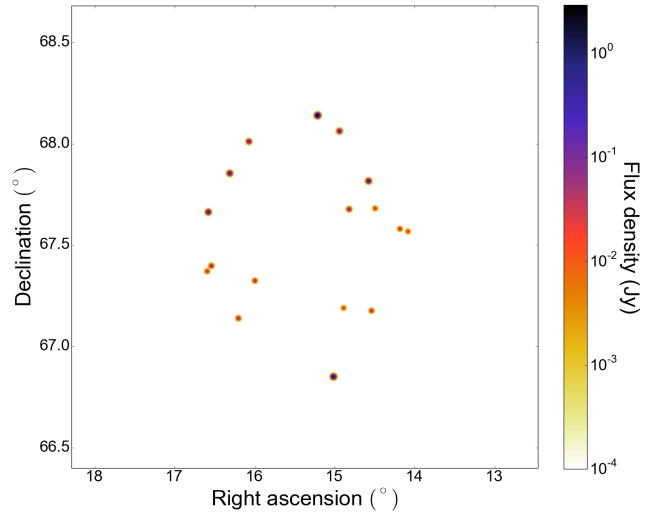}
 \caption{The simulated, noise-free sky model with 17 sources with flux densities varying between 0.03 and 3.13 Jy.}
\label{fig:field}
\end{figure}

\begin{figure}
\centering
\includegraphics[width=1\linewidth]{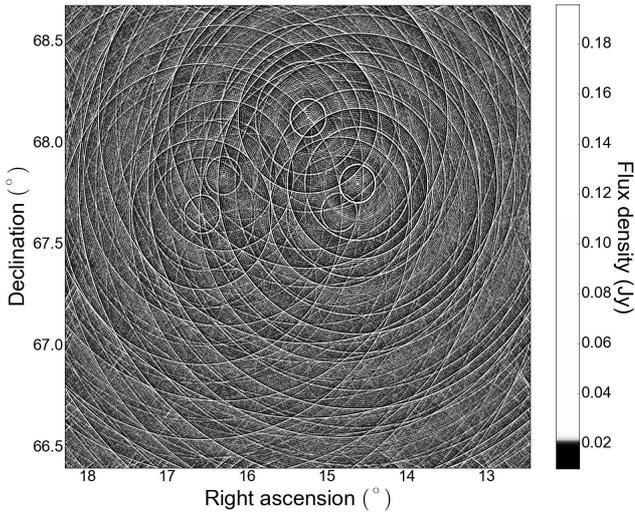}
 \caption{The dirty dataset for the model of Fig. \ref{fig:field}, as the telescope would see it (the colours are histogram-equalised to improve contrast). The image is produced directly from the visbilities and shows the typical ring structure around bright sources that is seen in interferometric data, due to the missing angular-scale information in the dataset. The rms noise in flux density is about 0.28 mJy.}
\label{fig:dirty}
\end{figure}

The simulated field consists of 17 unpolarised, point sources with known positions. The science goal was to determine the flux densities of these sources. We based the simulation on an existing field observed by WSRT, consisting of sources with a range of fluxes (from $0.03-3.13$ Jy). This is a very simple sky model, consisting only of point sources, whereas in the second example of the paper, we address modelling of extended sources. We do not explore the possibility of extended sources of arbitrary shapes, as this is out of the scope of this paper, but this should be possible using shapelets, such as employed in the existing PyBDSM  software\footnote{\label{note1}Python Blob Detection and Source Measurement software, \url{www.lofar.org/wiki/doku.php?id=public:user_software:pybdsm}}. The brightness matrix in Eq.(\ref{eq:rime_full}) for an unpolarised point source is written as:
\begin{equation}
\coh{B}{}^{\scriptscriptstyle \text{POINT}} =
\begin{pmatrix}
 I & 0\\
 0 & I
\end{pmatrix},
\end{equation}
where $I$ is the intensity.

\noindent Fig. \ref{fig:field} shows an image of the true input model without any instrumental effects, while Fig. \ref{fig:dirty} shows the dirty image of the sky.

 \subsubsection{Instrumental parameters}
{\bfseries Beam width}\\
Knowing the primary beam pattern is critical for any astronomical survey. Current practice is to determine the primary beam 
pattern using a technique such as holography \citep{scott}, then fix a beam model, without propagating any uncertainty information into the estimates of the science parameters. Since the primary beam directly attenuates the flux distribution of the sky, even a small error in the beam model can lead to large biases. We thus include beam parameters in our analysis. WSRT commonly adopts a simple model for the primary beam\footnotemark[1], namely: $\text{cos}^3(c\nu\theta)$, where $\nu$ is the observing frequency (in GHz), $\theta$ is the distance from the pointing centre in degrees and $c$ is the beam factor (in 1/GHz). The beam factor (or beam width) is known to vary slightly with frequency. As proof of concept, we assume it is unknown, and include it as a further instrumental parameter. One could provide a more complex model for the primary beam and easily fit those parameters with this technique as well, comparing the models with the Bayesian evidence. The model for the beam enters the RIME of Eq.(\ref{eq:rime_full}) as a direction-dependent Jones matrix:
\begin{equation}
\label{eq:jones_beam}
 \boldsymbol{E}^{\scriptscriptstyle \text{BEAM}}(l,m) = \text{cos}^3(c\nu\sqrt{l^2+m^2}) \,\boldsymbol{I},
\end{equation}
where $\boldsymbol{I}$ is the identity matrix.\\

\noindent{\bfseries Pointing errors}\\
Pointing errors can substantially corrupt radio observations and are known to be a limiting factor in deep observations with WSRT \citep{smirnov3} and other telescopes. The greatest effect is on sources on the flank of the primary beam,
where the gradient of the beam pattern is steep, and a small pointing error produces a larger error in apparent flux (compared
to the centre of the beam). Since the errors can be different from antenna to antenna, this produces errors on the observed 
visibility amplitudes, which translates into artefacts in the image. Essentially each source is `defocussed' in a complicated way.
Thus, we can immediately suspect there will be a correlation between the pointing errors and source flux densities. Two prior approaches
to inferring pointing errors directly from the data have hinged on maximum-likelihood estimates. These are the 
pointing selfcal algorithm \citep{pointing-selfcal} and direct fitting with MeqTrees (Smirnov 2011\footnote{\url{https://indico.skatelescope.org/getFile.py/access?contribId=20&sessionId=9&resId=0&materialId=0&confId=171}}). Neither approach
estimates the correlation between pointing errors and source parameters, which the Bayesian approach naturally provides. We inject time-varying polynomial pointing errors for each of the 14 WSRT antennas. We use a second order polynomial for each pointing error and fit for the coefficients. A polynomial pointing error in each orthogonal direction for each antenna results in a total of 84 pointing-error parameters. The pointing errors are written as a Jones matrix in Eq.(\ref{eq:rime_full}):
\begin{equation}
\label{eq:jones_pe}
 \boldsymbol{E}^{\scriptscriptstyle \text{PE}}_{p}(l,m)= \boldsymbol{E}^{\scriptscriptstyle \text{BEAM}}(l+\delta l_p,m+\delta m_p)\,
\end{equation}
where $\delta l_p$ and $\delta m_p$ are the pointing errors in the right ascension and declination direction respectively, for antenna $p$. The pointing errors are taken to be time-varying polynomials, written as:
\begin{equation}
 \delta l_p = c_2t^2+c_1t+c_0,
\end{equation}
and similarly for $\delta m_p$, where $t$ is time (rescaled over the observation) and $c_k$ are the coefficients we determine with MCMC.\\

\noindent{\bfseries Noise}\\
The noise on the visibilities is expected to be Gaussian, stationary and uncorrelated. Noise level can be estimated
with some precision from the known system temperature, here however we show than it can also be inferred accurately directly from the data.
We thus included one final parameter for the standard deviation of the noise on the visibilities. 

\subsubsection{Resulting measurement equation}
The RIME for this example problem is thus:
\begin{equation}
 \resizebox{0.99\linewidth}{!}{$\coh{V}{pq} = \displaystyle \sum_s \left( \boldsymbol{E}^{\scriptscriptstyle \text{BEAM}}(l_s+\delta l_p,m_s+\delta m_p) \boldsymbol{K}^{(s)}_p \coh{B}{s}^{\scriptscriptstyle \text{POINT}} \boldsymbol{K}_q^{(s),H} (\boldsymbol{E}^{\scriptscriptstyle \text{BEAM}})^H(l_s+\delta l_p,m_s+\delta m_p)\right),$}
 \label{eq:rime1}
\end{equation}

\noindent where $s$ runs from 1 to 17 over all the sources. 
This brings the total to 103 parameters: 17 scientific (the flux densities of the sources) and 86 instrumental (84 pointing error parameters, the beam width and the noise). The full model can be visualised in the Bayesian factor graph of Fig. \ref{fig:factor} and a more detailed description of factor graphs is given in section \ref{sec:factor} of the appendix.

\begin{figure}
\centering
\includegraphics[width=1\linewidth]{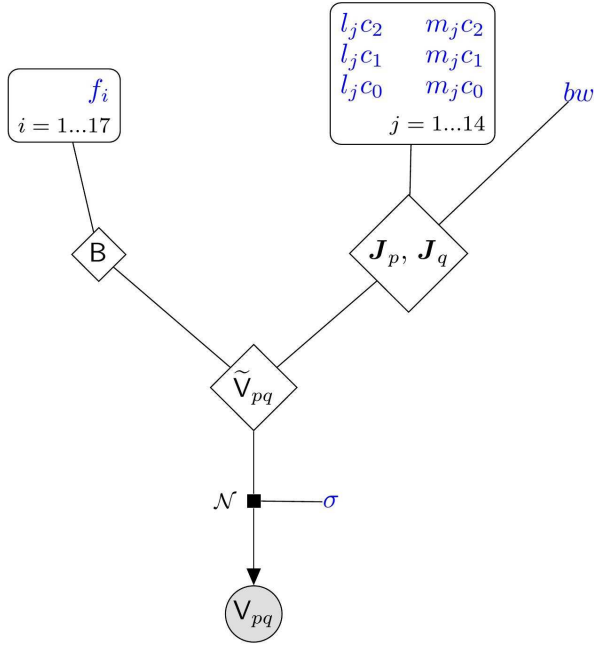}
\caption{Bayesian factor graph (see section \ref{sec:factor} of the appendix) of the model for the first simulated dataset. All parameters we estimate with MCMC are the constants, without any circles around them, coloured blue. The $\coh{V}{pq}$ are the observed visibilities, which are drawn from a normal distribution of mean $\coh{\widetilde{V}}{pq}$ (the unobserved, true visbilities) and standard deviation $\sigma$, which is one of the parameters we estimate with MCMC. These `true' visibilities are governed by the RIME, which is here simplified graphically to two components, the brightness matrix, $\coh{B}{}$, and the Jones' matrices of the antennas, $\boldsymbol{J}_p,\,\boldsymbol{J}_q$. The flux densities of the 17 sources are represented by $f_i$, which form components of $\coh{B}{}$. The coefficients of the polynomial time-varying pointing errors, $l_jc_k$ and $m_jc_k$ (where $j$ represents the antenna number and $k$ is the number of polynomial coefficient) enter the Jones matrices, along with the beam width, $bw$.}
 \label{fig:factor}
\end{figure}

\subsection{Using MCMC for joint parameter inference}
The initial step of our analysis was to choose an appropriate sky model in MeqTrees (specifying the brightness matrix in Eq.(\ref{eq:rime_full})) and select the telescope configuration corresponding to the dataset including all known sources of interference and instrumental errors (the Jones matrices in Eq.(\ref{eq:rime_full})). We vary all the parameters within the model -- the flux densities, pointing errors, beam width and noise -- using MCMC. Fig. \ref{fig:flowchart} illustrates how the sampling algorithm repeatedly calls MeqTrees with new parameter values and evaluates the likelihood.  MCMC uses the likelihood (Eq.(\ref{eq:likelihood})) to determine the best-fitting parameter values and to explore the surrounding parameter space, thus determining the uncertainties and correlations for all parameters.
 
\subsection{Technical details and priors}
Due to the large volume of the parameter space, we use a standard, gradient-based optimisation algorithm to get close to the best-fitting parameter values and provide a good starting point for the MCMC. We run several chains in parallel, each of around $500, 000$ steps, repeatedly computing and diagonalising the covariance matrix to improve convergence, and we test convergence using the Gelman-Rubin statistic \citep{gelman}. The estimated parameters and their uncertainties are determined by finding the mean and standard deviation (using percentiles) from the marginalised one-dimensional posterior for each parameter. For this particular setup, MeqTrees takes about 0.4s for one likelihood calculation, parallelised using 4 cores of 2.2 GHz each. As 10 chains were run, 40 cores in total were used resulting in approximately 55 CPU hours for convergence per dataset.

We apply a uniform prior to the pointing error parameters, restricting them to the broad range of $\pm200$ arcseconds.  We also restrict the beam width to be positive, and vary the noise on the visibilities in logarithmic space (with an infinitely broad prior in log-space). We do not restrict the ranges of the flux densities. 

\subsection{Comparison with CLEAN plus source extraction}
\label{sec:clean}
To compare our technique with the standard approach, we apply CLEAN followed by a source-extraction algorithm to determine the flux densities of the sources (we call this combination CLEAN+SE), without any instrumental calibration. We do not use any calibration algorithms such as self-cal, because it would have no benefit: our dataset only has direction-dependent instrumental effects, whereas self-cal can only correct for direction-independent effects. Current approaches to direction-dependent calibration are of no help here because:
\begin{enumerate}
 \item Direction-dependent solutions (such as peeling, or differential gains) can in principle solve for the variable gains induced by pointing error, given a prior source model. However, this destroys information on the source, since deviations between the true sky and the prior model are completely absorbed by such gain solutions.
 \item Pointing selfcal should in principle improve the CLEAN maps and thus produce better source model estimates. However, implementations of this remain unavailable to the public.
 \item MeqTrees should in principle be able to do a maximum-likelihood solution for the source parameters and pointing errors simultaneously. However, only solutions for the latter has been demonstrated to work in practice and as we have argued, a maximum-likelihood solution produces a point estimate for the parameters which may be biased due to correlations.
\end{enumerate}
Instead, we apply a na\"ive CLEAN algorithm, followed by source extraction, to compare with BIRO as a worst case scenario in the case of time-varying pointing errors. Note that we do provide prior information on the positions of the sources to CLEAN, in the form of CLEAN boxes.

We use the CLEAN implementation (specifically the Cotton-Schwab algorithm) in the software package CASA\footnote{Common Astronomy Software Applications, \url{http://casa.nrao.edu/}} to image the simulated datasets. The images were made with robust weighting with a robustness parameter of $-1.0$. We did 1000 iterations of CLEAN with a loop gain of 0.1. Interactive cleaning was performed on the visibility data twice, once with masks defined around known source positions and then with masks defined around only those sources that were found during the cleaning procedure.  The source extraction was performed interactively using PyBDSM to ensure that the artefacts were not wrongly identified as sources.

\subsection{Results}
\begin{figure}
\centering
\includegraphics[width=1\linewidth,trim=0cm 0cm 0cm 0cm, clip=true]{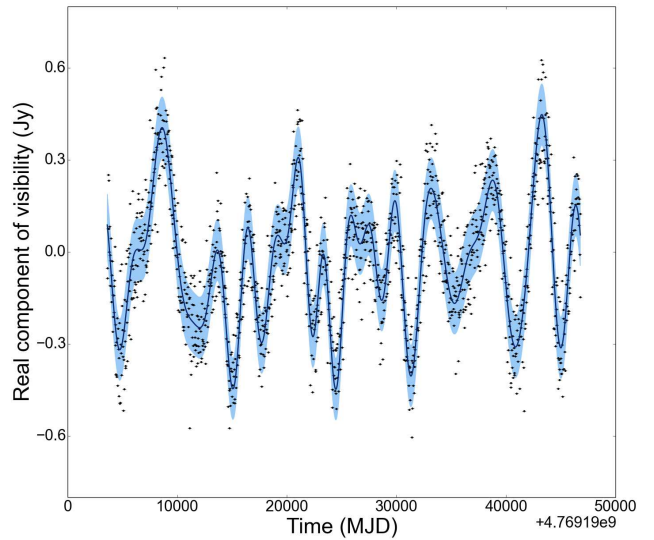}
 \caption{Example of fitting a model to the raw data. Plotted are the real component of the visibilities for a single baseline (between antenna 0 and 1) and for the single channel of the data, in black. The best fitting model line is overplotted in dark blue, with a band of uncertainty of 0.1 Jy (the original noise added to the simulation) in pale blue.}
\label{fig:vis}
\end{figure}

\begin{figure}
 \centering
 \includegraphics[width=0.95\linewidth]{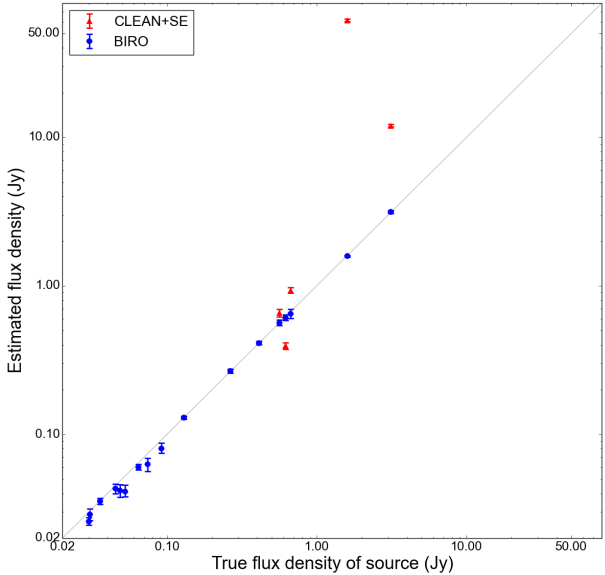}
\caption{Estimated vs true flux densities of the sources with error bars as estimated  by BIRO (blue circles) and by a CLEAN+Source Extraction algorithm (red triangles). Note that CLEAN+SE only detects 5 out of 17 sources. The BIRO error bars are the standard deviation of marginalised one-dimensional posterior for each flux parameter. While the BIRO results are unbiased, CLEAN+SE has two problems: it underestimates the error bars and yields biased estimates of the flux densities of up to $44\sigma$. The reader is reminded that this dataset contains no direction-independent effects that may normally cause biases in a CLEAN analysis; these biases are instead due entirely to the complexities in the dataset introduced by the time-varying pointing errors.}
\label{fig:compare}
\end{figure}

 To illustrate fitting a model to the raw data, we plot a subset of the visibilities in Fig. \ref{fig:vis} with the best-fit visibilities as obtained by BIRO. Fig. \ref{fig:compare} (with numerical details in Table \ref{tab:clean}) shows the comparison between the flux densities obtained by CLEAN+SE and those by BIRO. The flux densities of CLEAN+SE are on average biased due to undealt-with correlations with the pointing errors and underestimated uncertainties. Additionally, because of the time-varying pointing errors corrupting the data, CLEAN+SE only manages to find 5 of the 17 sources. With polynomial pointing errors included in the simulations, bright artefacts dominated the final image resulting in the weaker sources being swamped. In contrast, because these correlations are taken into account, the Bayesian approach is able to recover the true flux densities for all sources and to determine error bars that include the effects of all nuisance parameters. Without the instrumental errors, BIRO achieves similar flux estimates to CLEAN+SE.

\begin{figure*}
 \centering
\includegraphics[width=0.9\linewidth]{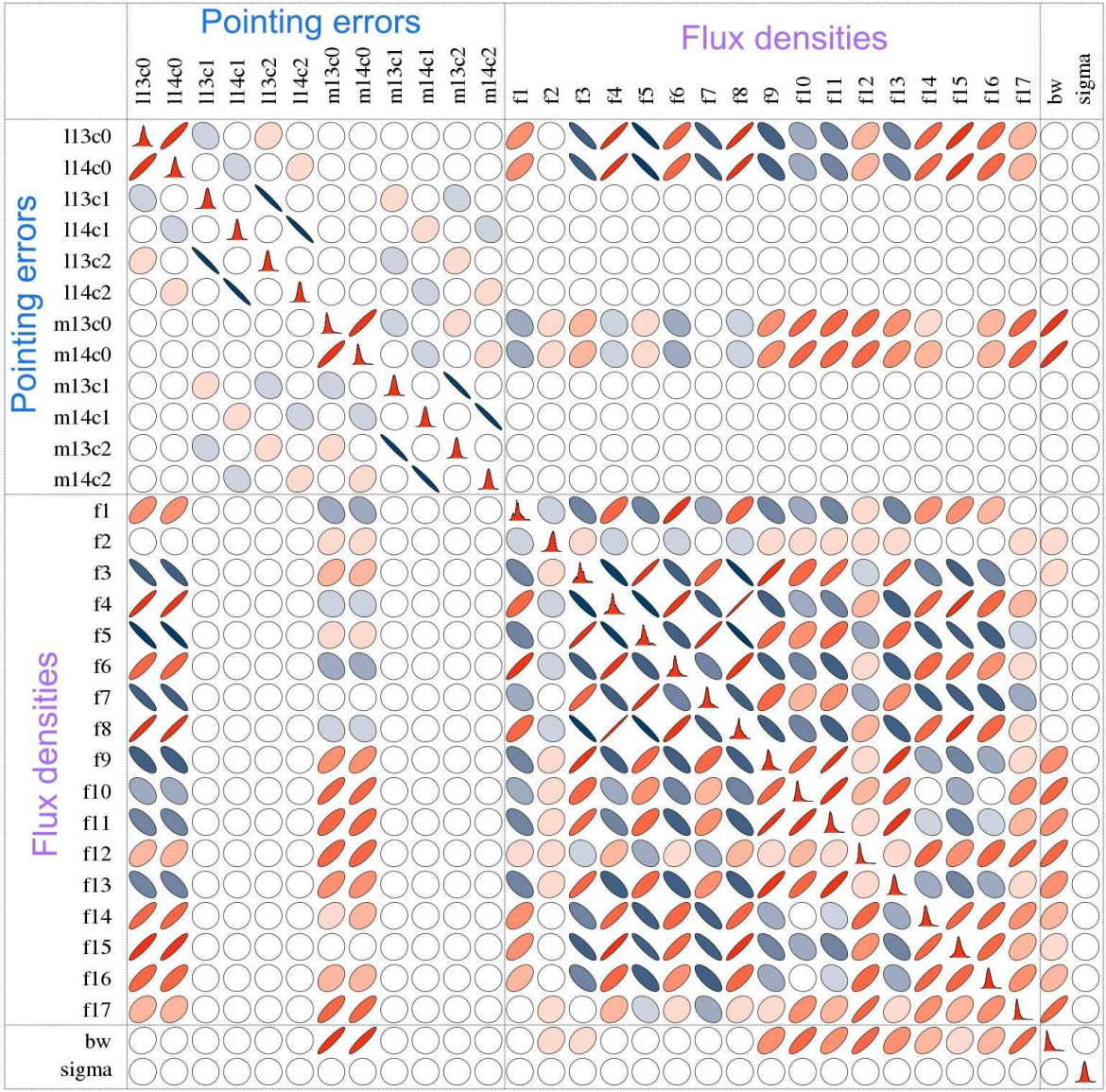}

\caption{Covariance matrix between a subset of parameters illustrating the strong correlations between the science and  instrumental parameters that must be accounted for to achieve unbiased results. The parameters are listed on each axis with the correlations between them represented by a coloured ellipse, either positive (red ellipse angled to right) or negative (blue ellipse angled to left). The leading diagonal shows the one-dimensional marginalised posterior for each parameter. For the pointing errors, $l_jc_k$ refers to the $k$'th coefficient of the polynomial time-varying pointing error in the right ascension direction for the $j$'th antenna and $m_jc_k$ is the same for the declination direction. The flux densities of the 17 sources are given by $f_i$, ordered from brightest to faintest, and $bw$ and $sigma$ represent the beam width and noise on the visibilities respectively.}
\label{fig:covmat}
\end{figure*}

\begin{figure}
  \centering
  \includegraphics[width=1\linewidth]{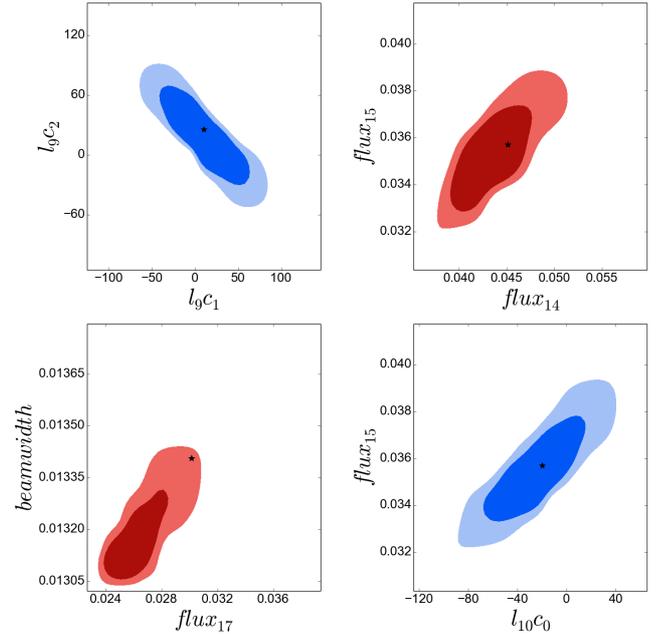}
  
  \caption{Credible interval contour plots between a subset of parameters. The 1$\sigma$ and 2$\sigma$ probability densities are shown in dark and light colours respectively. The true (input) parameters are marked with a black star. The pairs of parameters are: \emph{Upper left:} The two highest order coefficients of the pointing error in the right ascension direction for antenna 9. \emph{Upper right:} Flux densities of two of the sources. \emph{Lower left:} The flux density of the 17th source vs the beam width. \emph{Lower right:} The constant term from the polynomial pointing error in the right ascension direction for antenna 10 vs the flux density of the 15th source.}
  \label{fig:contour}
 \end{figure}

Fig.~\ref{fig:covmat} shows a subset of the covariance matrix between parameters and Fig. \ref{fig:contour} shows an example 1$\sigma$ and 2$\sigma$ contour plot between pairs of parameters. The key result of Fig. \ref{fig:covmat} is that it highlights the significant and complex correlations between the pointing errors and flux densities, i.e. the instrumental and science parameters, which therefore need to be estimated jointly allowing for the correlations. 

The (anti-)correlations between pointing errors and flux densities are easy to understand qualitatively. Consider a source on the flank of the main lobe of the primary beam, e.g. on the half-power point. If a given antenna mispoints \emph{towards} the source, the source will be subject to a higher primary beam gain, in other words, it will be perceived as brighter by all baselines involving that antenna. Mispointing \emph{away} from the source has the opposite effect. The nature of the correlation will also strongly depend on the position of the source with respect to the pointing centre. For example, a source near the centre of the main lobe (i.e. on a `flat' part of the primary beam pattern) will correlate very weakly with pointing error, while a source on the inner flank of the first sidelobe will correlate with mispointing \emph{away} rather than \emph{towards}. Since different baselines contribute to different Fourier mode measurements, pointing error will also have a complicated interaction with perceived source structure. Similar arguments apply to beamwidth. 

Deriving the exact quantitative nature of this correlation analytically is highly impractical, which is why a technique like BIRO proves so powerful. This covariance matrix could be used to assist in calibration, study calibration parameters or as input to future MCMC analyses on similar datasets.

 \begin{table*}
 \centering
 \renewcommand{\arraystretch}{1.4}
 \resizebox{0.9\linewidth}{!}{
  \begin{tabular}{*{6}{c}}
  \centering
  Name&RA ($^\circ$)&Dec ($^\circ$)&True Flux Density (Jy)&C+SE Flux Density (Jy)&BIRO Flux Density (Jy)\\ 
\hline 
F1&$15.216$&$68.139$&$3.1285$&$11.9723\pm{0.2240}\,(39.5\sigma)$&$3.1410^{+0.0828}_{-0.0710}\,(0.2\sigma)$ \\ 
F2&$15.017$&$66.849$&$1.6030$&$61.4220\pm{1.3557}\,(44.1\sigma)$&$1.5834^{+0.0185}_{-0.0185}\,(1.1\sigma)$ \\ 
F3&$16.583$&$67.662$&$0.6719$&$0.9345\pm{0.0415}\,(6.3\sigma)$&$0.6476^{+0.0478}_{-0.0410}\,(0.5\sigma)$ \\ 
F4&$14.578$&$67.815$&$0.6170$&$0.3942\pm{0.0191}\,(11.7\sigma)$&$0.6114^{+0.0249}_{-0.0249}\,(0.2\sigma)$ \\ 
F5&$16.315$&$67.853$&$0.5648$&$0.6579\pm{0.0372}\,(2.5\sigma)$&$0.5635^{+0.0272}_{-0.0227}\,(0.0\sigma)$ \\ 
F6&$14.943$&$68.061$&$0.4115$&$Not\, found$&$0.4127^{+0.0135}_{-0.0112}\,(0.1\sigma)$ \\ 
F7&$16.075$&$68.010$&$0.2640$&$Not\, found$&$0.2671^{+0.0093}_{-0.0093}\,(0.3\sigma)$ \\ 
F8&$14.822$&$67.676$&$0.1293$&$Not\, found$&$0.1295^{+0.0031}_{-0.0031}\,(0.1\sigma)$ \\ 
F9&$16.546$&$67.396$&$0.0919$&$Not\, found$&$0.0805^{+0.0071}_{-0.0059}\,(1.6\sigma)$ \\ 
F10&$16.207$&$67.138$&$0.0742$&$Not\, found$&$0.0629^{+0.0064}_{-0.0064}\,(1.8\sigma)$ \\ 
F11&$16.001$&$67.323$&$0.0645$&$Not\, found$&$0.0599^{+0.0029}_{-0.0023}\,(1.6\sigma)$ \\ 
F12&$14.541$&$67.175$&$0.0524$&$Not\, found$&$0.0413^{+0.0041}_{-0.0031}\,(2.7\sigma)$ \\ 
F13&$16.598$&$67.370$&$0.0487$&$Not\, found$&$0.0419^{+0.0040}_{-0.0040}\,(1.7\sigma)$ \\ 
F14&$14.186$&$67.579$&$0.0451$&$Not\, found$&$0.0430^{+0.0031}_{-0.0031}\,(0.7\sigma)$ \\ 
F15&$14.496$&$67.680$&$0.0357$&$Not\, found$&$0.0354^{+0.0017}_{-0.0015}\,(0.2\sigma)$ \\ 
F16&$14.085$&$67.566$&$0.0306$&$Not\, found$&$0.0289^{+0.0026}_{-0.0026}\,(0.6\sigma)$ \\ 
F17&$14.891$&$67.188$&$0.0301$&$Not\, found$&$0.0259^{+0.0017}_{-0.0013}\,(2.5\sigma)$ \\ 

   \end{tabular}}
\caption{Comparison between the CLEAN+source extraction results (shortened to C+SE) and the BIRO results for the flux densities (in Jy) of the sources in the dataset. The bias in terms of number of standard deviations away from the true flux density is given in brackets. For the five sources CLEAN+SE found, the error on the position was less than $10^{-4}$ degrees.}
\label{tab:clean}
 \end{table*}

\section{Example 2: Model comparison}
\label{sec:source_sep}
In this example problem, we show that BIRO is able, using model selection \citep{jeffreys,trotta}, to choose the correct model in each of three different cases, distinguishing between an extended source, an unresolved point source and two close (sub-synthesised-beam) sources. The sources recovered are all smaller than the synthesised beam. This is known as super-resolution and has recently been shown to be possible with compressive sensing \citep{wiaux,li,carrillo1,carrillo2,honma} (and to some extent \cite{marti}). Here we use the Bayesian evidence to determine the correct model of these sub-synthesised-beam sources, with statistical significance. Although in this example problem we exclude instrumental effects, they can, in general, be included as in example 1.

\subsection{Simulated datasets and models}
The datasets for this example use the same frequency, bandwidth, integration time and noise characteristics as the dataset simulated in section \ref{sec:joint}. We simulate three datasets with three different sky models with all the sources away from the phase centre:
a point source, a sub-synthesised-beam extended source modelled as a Gaussian and two point sources separated by the distance the size of that Gaussian. No instrumental effects were included in the model-selection simulations and the beam width and noise were assumed to be known. Fig. \ref{fig:source_sep} shows the input model for all three cases in the left column.

\begin{figure}
  \centering
  \includegraphics[width=0.6\linewidth]{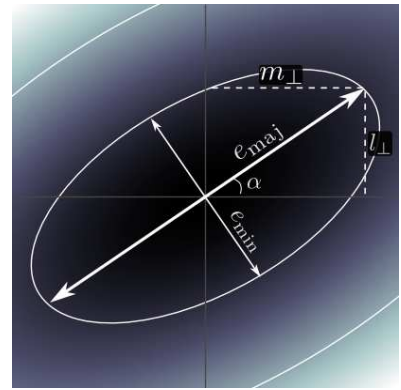}
  \caption{The parameterisation of a Gaussian extended source in MeqTrees. Here, $e_{\text{maj}}$ and $e_{\text{min}}$ are the major and minor axes of the Gaussian and $\alpha$ is the position angle. MeqTrees uses $l_{{\scriptscriptstyle \perp}}$, $m_{{\scriptscriptstyle \perp}}$ and $r= e_{\text{min}}/e_{\text{maj}}$ in its parameterisation of a Gaussian.}
  \label{fig:gaussian}
 \end{figure}

The point sources are parametrized by the Stokes I flux density and the position as the distance from the phase centre, along two mutually perpendicular axes, $l$ and $m$. The extended Gaussian source has three more parameters in the form of the projections of the major axis on the $l$ and $m$ axes and the ratio of the minor to major axis, defined as:
\begin{gather}
\label{eq:gaussshape}
 l_{{\scriptscriptstyle \perp}}=e_{\text{maj}}\,\text{sin}(\alpha)\\
 m_{{\scriptscriptstyle \perp}}=e_{\text{maj}}\,\text{cos}(\alpha)\\
 r= e_{\text{min}}/e_{\text{maj}},
 \end{gather}
 where $e_{\text{maj}}$ and $e_{\text{min}}$ are the major and minor axes of the Gaussian source and $\alpha$ is the position angle (the angle of rotation of the extended source). See Fig. \ref{fig:gaussian} for a visual description. The brightness matrix of Eq.(\ref{eq:rime_full}) for an extended Gaussian is simply the product of a Gaussian and the brightness matrix for a point source. The RIME is simple in this example, since there are no instrumental effects apart from the usual phase shift between antennas:
 \begin{equation}
 \coh{V}{pq} = \displaystyle \sum_s \left( \int \int_{lm} \boldsymbol{K}^{(s)}_p f(l,m) \coh{B}{s}^{\scriptscriptstyle \text{POINT}} \boldsymbol{K}_q^{(s),H} \text{d}l \text{d}m \right),
\end{equation}
\noindent where $f(l,m)$ is a Gaussian in $l$ and $m$ for the extended source case and $f$ is a delta function for the one and two-source models. Also in the one and two-source models, $l$ and $m$ reduce to single points $l_s$ and $m_s$, as in Eq.(\ref{eq:rime1}).
 
\subsection{Using MultiNest for model selection}
We use MultiNest for calculating the Bayesian evidence (see section \ref{sec:bayes}) and MeqTrees for predicting the model visibilities from the sampled
source parameters from which the likelihood is computed iteratively. The likelihood is computed according to Eq.(\ref{eq:likelihood}).
The posterior probability distributions are obtained as a by-product along with the uncertainties in the best-fit parameter values and the Bayesian evidence.

For the single-point-source model, we vary three parameters: the flux density and relative source position, $l$ and $m$. We similarly vary the flux densities and positions of the two sources in the two-source model. The Gaussian extended source model has six parameters: the flux density, position coordinates, and the shape parameters ($l_{{\scriptscriptstyle \perp}}$, $m_{{\scriptscriptstyle \perp}}$, $r$).

We generate a unique, simulated dataset for each of the three cases and then fit each of the three models to them, to see if the correct model is selected in each case. MultiNest fits for the parameters, their uncertainties and correlations (just as MCMC does in example 1), but also returns the evidence, $\mathcal{Z}\left(\mathrm{\mathbf{D}}| H\right)$ (the probability of the data, given the hypothesis). By taking the ratio of evidences, one can determine whether one model is favoured over another, and by how much. The Jeffrey's scale \citep{jeffreys,trotta} provides an intuitive way of deciding whether the evidence is strong enough to select a model, based on odds derived directly from the evidence. 

\subsection{Technical details and priors}
We use uniform priors for all the source parameters. The flux density is restricted to the range 0 to 2 Jy. The position parameters are allowed to be both positive and negative in the range -$25\arcsec$ to $25\arcsec$ since the position is measured relative to the phase centre. For the shape parameters of the extended source, ($l_{{\scriptscriptstyle \perp}}$ and $m_{{\scriptscriptstyle \perp}}$), we allow the prior ranges to be big enough to encompass the point-spread-function (PSF) of the interferometer and no more, since we are dealing with sub-synthesised-beam sources. This translates to a range of $0\arcsec$ to $20\arcsec$ for $l_{{\scriptscriptstyle \perp}}$ and -$20\arcsec$ to $20\arcsec$ for $m_{{\scriptscriptstyle \perp}}$. Finally, we restrict the minor-to-major axis ratio ($r$) to be positive, but less than unity to be physically meaningful.
We found that using 1000 live points achieved good results from MultiNest.

\subsection{Results}
 The relative logarithmic evidences are computed for each model giving the relative confidence with which one model is preferred over another (see Table \ref{tab:evmatrix}). We find that the correct hypothesis is selected in all cases, at odds of $10^{593}$:1, $10^{993}$:1 and 62:1, for the two-point-source, extended-source and single-point-source models respectively. Using model selection, BIRO is able to select the correct model in all three cases (the model with the highest evidence), showing it can perform source separation even on sub-synthesised-beam scales.
 
\begin{figure}
 \centering
\includegraphics[width=1\linewidth]{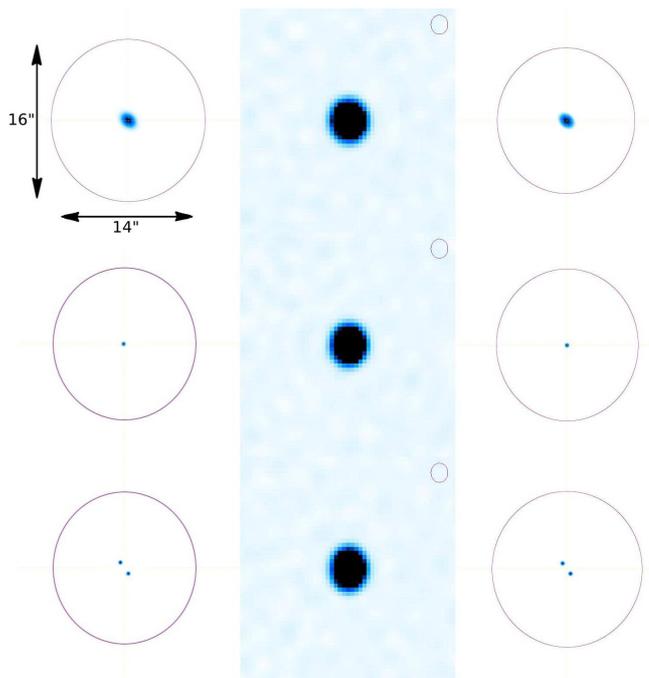}
\caption{\emph{Left column:} The true sky for the extended Gaussian, single point source and two point sources models (from top to bottom). \emph{Middle column:} The CLEANed image for the three models. \emph{Right column:} The maximum posterior BIRO image for the three models. The purple contour in each image indicates the size of the synthesised beam, as returned by CLEAN (note that the sources are all much smaller than the synthesised beam). BIRO recovers the correct input model each time while CLEAN is unable to distinguish between the models at the same SNR (in this case the SNR was 1000).}
\label{fig:source_sep}
\end{figure}

 We computed a `best-fitting' image by running MeqTrees with the maximum posterior model and parameters in each of the three cases, to compare with the CLEANed image (see Fig. \ref{fig:source_sep}). We use the same CLEAN parameters as in section \ref{sec:clean}. The CLEANed images (at least in this case, without an enforced smaller beam size) are unable to reach the sub-synthesised beam scales achievable by BIRO.
 
\begin{figure}
  \centering
  \includegraphics[width=0.99\linewidth]{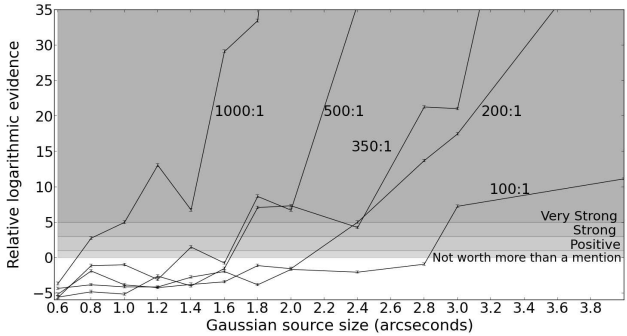}
  \caption{Relative natural log-evidence (i.e. the natural logarithm of the ratio of the Bayesian evidence for the true model to that of a single point source) as a function of Gaussian source size, for the extended source input model, showing the evidence-crossover points for different source sizes and signal-to-noise ratios (peak flux to background noise). The horizontal axis gives the size of the circular Gaussian source in the input model (the reader is reminded that the FWHM of the synthesised beam is around 13 arcseconds). The vertical axis gives the odds in favour of the Gaussian source model when model comparison is performed for the Gaussian model against a point-source model. The more positive the relative log-evidence is, the more strongly is the Gaussian model favoured. Each curve on the graph is for a different noise level with the approximate (map) SNRs shown in the legend.}
  \label{fig:crossover}
  \end{figure}

 In Fig. \ref{fig:crossover}, we determine the point at which model selection fails to distinguish an extended source from a point source for different source sizes and signal-to-noise ratios (SNRs). Any evidence lower than `strong' is not usually considered high enough to say either way which model is correct. Perhaps obviously, at high SNR extremely small sources can be detected (around 1.0 arcseconds) and sources become more difficult to distinguish as the SNR is reduced. 
 
 Video \ref{vid:gaussian} in the online-only content shows visually how MultiNest converges to the correct model, exploring the posterior as it goes, for the extended source model. Each frame is an image generated using the parameters from every $40^{\text{th}}$ step of the chain. 

\begin{table}
\centering
\renewcommand{\arraystretch}{1.2}
\resizebox{1.\linewidth}{!}{
\centering
\begin{tabular}{c|ccc}
\hline
&\multicolumn{3}{c}{Hypothesis} \\
\cline{1-4}
Simulation    & A & B & C \\
\hline
A      & \hspace{28pt}$1:1$    & \hspace{9pt}$10^{593}:1$ & $10^{7200}:1$      \\
B      & \hspace{10pt}$10^{993}:1$     & \hspace{28pt}$1:1$ & $10^{5079}:1$      \\
C      & \hspace{22pt}$62:1$     & \hspace{18pt}$857:1$ & \hspace{21pt}$1:1$      \\
\hline
\end{tabular}
}
\caption{Relative evidences for each model in each simulated dataset. A is the two-source model, B is the extended source model and C is the one-source model. The evidences are relative to the model used to generate the dataset (so, for example, for the two-point-source dataset, the evidence for each model is compared to the two point source model). The maximum error in log-evidence is 1.5. High odds indicate the input model is favoured (as it is in all three cases), showing that nested sampling selects the correct model at high significance (at a SNR of 1000).}
\label{tab:evmatrix}
\end{table}

\section{Discussion and conclusions}
\label{sec:conclusions}
We have introduced the technique Bayesian Inference for Radio Observations (BIRO), a Bayesian approach to the deconvolution problem of radio interferometry. Instead of making an image and then performing source extraction, BIRO uses MCMC or nested sampling to fit models directly to the visibility data and obtain the posterior for the parameters of interest, as well as nuisance parameters.

In the first example problem, we focused on the relationship between scientific and instrumental parameters. It was found that all parameter estimates from BIRO were consistent within their error bars with the true values. As well as determining the uncertainties of the parameters, BIRO also returns the covariance matrix between them, as a by-product of the full posterior. Our work shows these correlations are complicated and non-negligible. BIRO effortlessly incorporates the effects of the correlations in the estimates of the marginalised uncertainties on the individual parameters, as well as providing a way to study these correlations in the form of the covariance matrix. We compared our results to a standard CLEAN algorithm, without calibration (since our simulated data contains \emph{only} direction-dependent effects and publicly available calibration algorithms only deal with direction-independent effects). Because of the time-varying pointing errors we introduce to the dataset, CLEAN is only able to find 5 out of the 17 sources and returns biased flux densities for them, while BIRO returns unbiased flux densities for all sources. BIRO is also able to correctly determine the coefficients of the time-varying pointing errors, the primary beam width and the noise on the visibilities.

In the second example problem, we addressed the issue of how to determine the best sky model for the data. We worked with three models: a single point source, a Gaussian extended source and two point sources. We simulated data for each of the three models and then, for each dataset, ran MultiNest to fit each of the three models and determine the Bayesian evidence. The evidence then determines the selection of the correct model. All of the sources detected were several times smaller than the synthesised beam, hence we successfully achieved super-resolution as well as source-separation.

This paper constitutes a proof of concept but more work is required before the technique can be easily applied to interferometric dataset:

\begin{enumerate}
\item Firstly, while using a WSRT simulation has relevance to the SKA due to the similar instrumental setup, the SKA will have many more antennas (on the order of a thousand) which will of course result in many more instrumental parameters (and indirectly more science parameters as the source count increases with sensitivity). Fortunately, while the number of instrumental parameters scales as the number of antennas, $N$, the number of datapoints scales as the number of baselines, i.e. ${\cal O}$($N^2$), meaning it is plausible that one could simultaneously determine the sky and instrumental parameters for large $N$. While the precedent for sampling an extremely large parameter space exists \citep{jasche}, new and sophisticated sampling techniques \citep{hmc1,hmc2,affine1,affine2} (which are also easily parallelised) will be required to improve convergence in the thousand-parameter regime, especially as the non-linear nature of the modelling makes sampling inefficient (as addressed in \cite{jasche}).

\item Secondly, the Bayesian approach is far more computationally intensive than standard deconvolution, taking hours (55 CPU hours in the case of example 1) to converge to the correct posterior distribution. The complexity of the likelihood computation scales as the number of antennas squared (i.e. the number of baselines), making an SKA-like computation difficult with the current setup. However, the RIME is intrinsically highly parallelisable allowing an efficient implementation of MeqTrees on GPUs. Preliminary work on a GPU implementation indicates a speed-up of the likelihood computation of about 250 times \citep{perkins}. This means this technique can be applied to data from existing telescopes such as ALMA \citep{alma} and LOFAR \citep{lofar}, using current computer clusters.

\item Thirdly, we need to address the problem of not knowing the sky model beforehand, which is a common difficulty when dealing with calibration but is particularly important here, as a Bayesian analysis relies on a good model. There are a number of ways to tackle this issue which we hope to address in future publications. A simple, but computationally-intensive, solution would be to run several different models (with increasing numbers of sources) and select between them using the Bayesian evidence. Another possible approach is to use a deconvolution algorithm, like CLEAN or RESOLVE, to get an initial set of sources and then iterate between deconvolution and the best fit of BIRO to get a subsequently better model. A more rigorous solution would be to use an algorithm like birth-death \citep{stephens} or reversible jump \citep{green} MCMC, which is able to determine both the number of parameters required and the posterior for them simultaneously. A further possibility is to combine the more general approach proposed in \cite{sutter2} and \cite{junklewitz2}, that divides up the field into many `pixels' that are then allowed to vary, with the calibration capabilities of BIRO to produce estimates of the sky model. This is even more computationally challenging however but would provide a more general and robust solution.\\
\end{enumerate}
BIRO is not only useful for dealing with systematics, which will become more important as telescopes become more sensitive, but it is also a powerful technique for lending statistical strength to topical scientific questions. Potential applications include: structures of black hole systems, jet emission in active galaxies, time variability of objects and radio weak lensing. BIRO allows a holistic way to include instrumental effects while at the same time returning the science we are interested in. By leveraging the power of Bayesian statistics, BIRO uses all information available to get the most out of interferometric datasets.

  \begin{video}
  \caption{Online only (also available at \url{https://vimeo.com/117391380}). Images generated from the MultiNest chain for the extended Gaussian source dataset and model. At every 40th step in the chain, that step's parameters were used to generate an image of the field. The parameters are at first quite variable but soon converge to the correct shape, position and flux density for the source. The sample probability, which is the normalised posterior for that point, improves as the chain converges to the correct parameter values.} 
  \label{vid:gaussian}
 \end{video}

\section*{Acknowledgements}
We would like to thank Paul Sutter for his useful referee comments. M.L. and J.Z. are grateful to the South Africa National Research Foundation Square Kilometre Array Project for financial support. M.L. acknowledges support from the University of Cape Town and resources from the African Institute for Mathematical Sciences. O.S. is supported by the South African Research Chairs Initiative of the Department of Science and Technology and National Research Foundation. B.B. acknowledges funding from the South African National Research Foundation. I.N. acknowledges the MeerKAT HPC for Radio Astronomy Programme. Part of the computations were performed using facilities provided by the University of Cape Town's ICTS High Performance Computing team: \url{http://hpc.uct.ac.za}. This project was initiated at the SuperJEDI Mauritius conference.

\section*{Author Contributions} 
M.L. performed the research for the joint estimation example and wrote the majority of the paper. I.N. performed the research for the source separation problem and wrote the corresponding sections of the paper and did some CLEAN analysis and wrote the corresponding section. J.Z. assisted with the source separation research and wrote part of the paper. O.S. provided the simulated datasets and led the technical aspects of radio observation modeling. B.B. developed the original idea. N.O. performed some CLEAN analysis. M.K. contributed to the conceptual development of ideas and provided support with the statistical methods. All authors commented on the research and edited the paper.

\label{References}
\footnotesize{
\bibliographystyle{mn2e}
\bibliography{refs}}

\appendix
\section{Bayesian Factor Graphs}
\label{sec:factor}
\begin{figure}[H]
 \centering
 \includegraphics[width=0.3\linewidth]{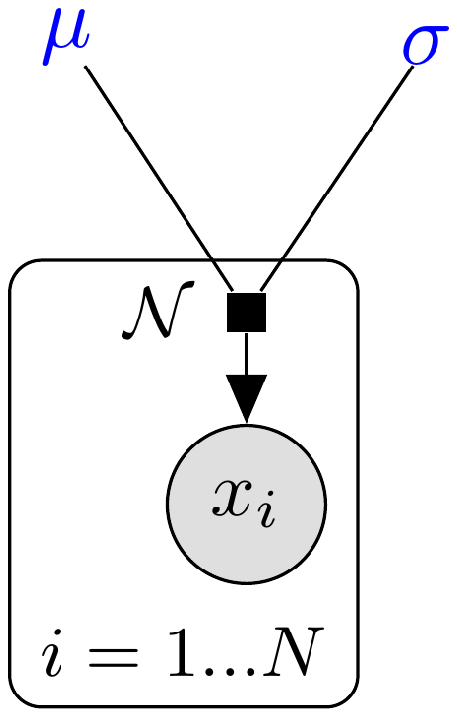}
 \caption{A simple example factor graph. In this model, the data are represented by a vector $x_i$, which we suspect is normally distributed. This is modelled by a normal distribution (represented by the factor labeled $\mathcal{N}$) which is governed by the parameters $\mu$ and $\sigma$. These constants would be the parameters we would want to estimate with an MCMC or MultiNest analysis.}
 \label{fig:factor_ex}
\end{figure}

Here we introduce Bayesian factor graphs, useful tools for visualising Bayesian models, which we use to describe the model of section \ref{sec:joint}. We make use of the directed factor graph notation, developed in \cite{dietz}, to visualise how the parameters in our models depend on one another. Table \ref{table:factor} defines the graphical primitives of a factor graph. Figure \ref{fig:factor_ex} demonstrates the use of the factor graph notation in a simple example.

\begin{table*}
  \caption[Factor graph node types]{Factor graph node types (adapted from \cite{dietz}). The concept of a plate is worth an extra mention. Frequently in models variables are repeated, such as the 17 flux densities or 14 sets of pointing errors in our model in example 1. A plate in a factor graph allows one to easily show these variables are repeated, but each can have a unique value. So in the case of the source flux densities, $m$ would range from 1 to 17, the value of $\mathcal{M}$.}
  \centering
  \includegraphics[width=0.7\linewidth]{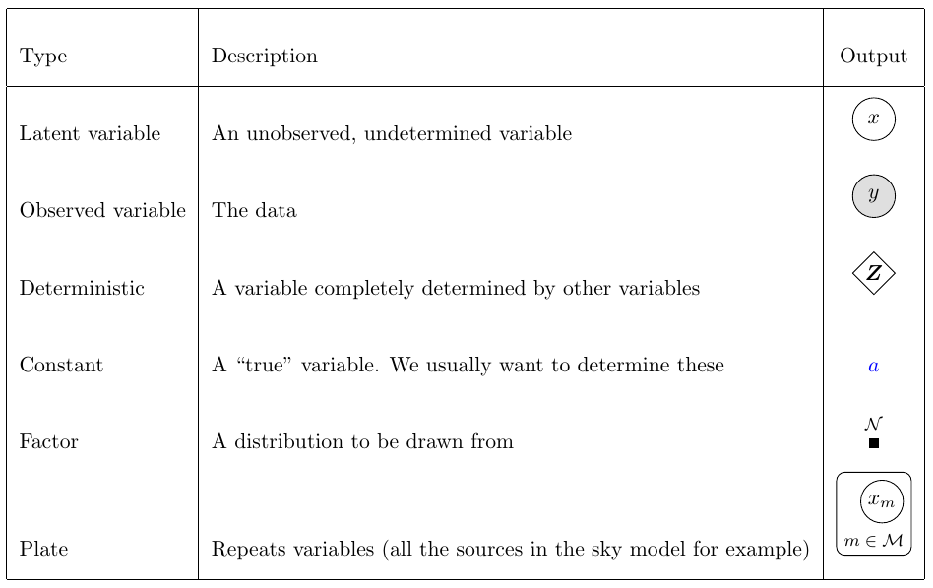}
  \label{table:factor}

\end{table*}

\label{lastpage}

\end{document}